\documentstyle[aps,graphics,epsfig]{revtex}
\begin{document}
\draft
\title{Quantum phase transitions in a linear ion trap}
\author{G.J.Milburn~$^*$ and Paul Alsing$^\dagger$}
\address{$^*$~Centre for Quantum Computer Technology,\\
and Department of Physics, The University of Queensland,QLD 4072 Australia.\\
$^\dagger$~The University of New Mexico\\
Albuquerque High Performance Computing Center\\
Albuquerque, New Mexico 87131}
\date{\today}
\maketitle

\begin{abstract}
We show that the quantum phase transition of the Tavis-Cummings model can be realised in a linear ion trap of the kind
proposed for quantum computation. The Tavis-Cummings model describes the interaction between a bosonic degree of freedom and a
collective spin. In an ion trap, the collective spin system is a symmetrised state of the internal electronic states of N
ions, while the bosonic system is the vibrational degree of freedom of the centre of mass mode for the ions.    
\end{abstract}

\section{Introduction}
More than two decades ago, when quantum optics was young,  the quantum dynamics of
collective spin systems interacting with a single bosonic degree of freedom was a major research problem. The
model arose as an attempt to describe the interaction between a collection of two level atoms and a single mode of the
radiation field.  Walls and co workers\cite{Collective}  were among the first to realise that such models provided ideal
examples of the role of quantum fluctuations in the nonlinear interaction between matter and light. Quantum fluctuations were
shown to drastically change the predictions of semiclassical theory in such systems. This phenomenon has appeared more recently
in the discovery of quantum phase transitions in quantum spin glasses\cite{Reiger} and other many body
quantum systems.  While the collective spin models did not directly apply to achievable experiments at the time, they did
provide insight that subsequently proved important for many other quantum optical experiments including
anti-bunching, squeezing\cite{WallsMilb}, and cavity QED\cite{Kimble}. In this paper we show that the
models of a collective spin interacting with one or more bosonic modes can now be experimentally realised in modern ion trap
systems of the kind proposed for quantum computation\cite{NIST,LANL}.  An enormous effort has gone into making such systems
work at the quantum level, with little interference form classical sources of noise, and a number of such experiments exist
today. It would thus appear worthwhile to  reconsider the collective spin models, and  the associated quantum many-body
effects exhibited by such systems, with a view to direct experimental realisation.

In particular we consider the Tavis-Cummings (TC) model\cite{tavis-cum}, which can be realised in a linear ion trap of $N$
ions with the bosonic degree of freedom appearing as the quantised collective centre-of-mass motion. If each ion is coupled to
the vibrational motion using an identical external (classical) laser  detuned to the first red-sideband transition, the
symmetry is such that the electronic degree of freedom for the ions can be described as a collective spin ($N$) and the
reversible dynamics is well described by the TC model. The TC model is known to exhibit important nonlinear quantum effects
including a quantum phase transition\cite{Reiger} in which the (zero temperature)  ground state undergoes a morphological
change as a parameter is varied and averages of intensive quantities undergo a bifurcation.

\section{The Tavis-Cummings model}
The interaction Hamiltonian for N ions interacting with the centre of mass vibrational mode can be controlled by using
different kinds of Raman laser pulses. A considerable variety of interactions has already been achieved
or proposed \cite{NIST,LANL,James}. In this paper we consider the first red-sideband transition. The ion is assumed to
be in a three dimensional anisotropic harmonic potential. Two dimensions are very tightly bound and are neglected. In the
remaining dimension, an external laser  couples the electronic state to the vibrational motion. If the vibrational
frequency is large enough and the Lamb-Dicke limit\cite{NIST} applies the motional sidebands of the absorption of the
electronic transition can be resolved and a laser detuned below the electronic resonance by one unit of the trap frequency can
excite the electronic transition by absorbing one vibrational phonon, the additional energy required being made up by the
laser. We will assume that the laser ( or lasers if a Raman process is used)  is sufficeintly strong that it can be treated
classically. Under these assumptions the Hamiltonian, in the interaction picture, is
\begin{equation}
 H_I=\hbar\Omega\sum_{i=1}^N(a\sigma_+^{(i)}+a^\dagger \sigma_-^{(i)}) 
\end{equation}
where the coupling constant is
$ 
\Omega=\eta\Omega_0
$
where $\eta^2=E_r/(\hbar M\omega_0)$ is the Lambe-Dicke parameter with $E_r$ the recoil kinetic energy of the atom, $\omega_0$
is the trap vibrational frequency, and $M$ is the effective mass for the centre-of-mass mode. The Lamb-Dicke limit assumes
$\eta\ <<\  1$, which is easily achieved in practice. The frequency, $\Omega_0$ is the effective Rabi frequency for the
electronic transition involved. The raising and lowering operators for each ion are defined by $\sigma_-=|g\rangle
\langle e|$ and
$\sigma_+=|e\rangle\langle g|$. This sideband transition can be used to efficiently cool the ions to the collective
centre-of-mass ground state, thus preparing the system in the vibrational ground state\cite{NIST}. 

If the external laser field on each ion is identical (in amplitude and phase) the interaction Hamiltonian is 
\begin{equation}
H_I=\hbar\Omega(a\hat{J}_+ +a^\dagger \hat{J}_-)
\label{TC}
\end{equation}
where we have introduced the bosonic annihilation operator $a$ for the centre-of-mass vibrational mode and 
where we have used  the definition of the collective spin operators,
\begin{equation}
\hat{J}_\alpha=\sum_{i=1}^N\sigma^{(i)}_\alpha
\end{equation}
where $\alpha=x,y,z$ .  Identical laser
fields could easily be obtained by splitting a single, stabilised laser into multiple beams.  The interaction Hamiltonian in
Eq (\ref{TC}) specifies the Tavis-Cummings model\cite{tavis-cum}. This model first appeared in quantum optics where the bosonic
mode is the quantised field in a cavity. However this realisation is difficult to achieve experimentally. In contrast the
vibrational mode realisation should be readily achieved. The dynamics resulting from this Hamiltonian is quite rich.  
Collective spin models of this kind were considered many decades ago in quantum optics\cite{Drummond,Dicke}. In much of that
work however the collective spin underwent an irreversible decay. In the case of an ion trap model however we can neglect such
decays due to the long lifetimes of the excited states. On the other hand heating of the vibrational centre-of-mass mode can
induce irreversible dynamics in the system in a manner that has not been previously considered, and that is reminiscent of
thermal effects in condensed matter physics.

We are interested in the driven Tavis-Cummings
model in which the vibrational mode is subject to a linear forcing term which can easily be achieved by a suitable
combination of Raman laser pulses, or by appropriate  AC voltages applied to the trap electrodes\cite{NIST}. In this case the
Hamiltonian, in the interaction picture, is given by 
\begin{equation}
 H_I=\hbar\Omega(a\hat{J}_+ +a^\dagger \hat{J}_-)+\hbar E(a+a^\dagger)
\label{drivenTC}
\end{equation}
This may be written in terms of the hermitian canonical oscillator variables $\hat{X}=(a+a^\dagger)/\sqrt{2}$, 
$\hat{Y}=-i(a-a^\dagger)/\sqrt{2}$, and the canonical angular momentum variables $\hat{J}_x=(\hat{J}_++\hat{J}_-)/2$,
$\hat{J}_y=-i(\hat{J}_+-\hat{J}_-)/2$, 
$\hat{J}_z=[\hat{J}_+,\hat{J}_-]/2$. It takes the form
\begin{equation}
H=\hat{X}\hat{J}_x-\hat{Y}\hat{J}_y+\chi\hat{X}
\end{equation}
with $\chi=E/\Omega$ and we have scaled the Hamiltonian by $H\rightarrow H/\sqrt{2}\Omega$. This indicates that time is
measured in units of $\frac{1}{\sqrt{2}\Omega}$.  

Alsing\cite{Alsing} has shown that the ground state of this system, for weak driving, is a product state in which the bosonic
mode is squeezed and the electronic states are rotated in the angular momentum space. We provide a direct proof of this
statement below. However it is first useful to consider the dynamics of the equivalent semiclassical model as many of the
results in the quantum case can be interpreted in terms of the features of the semiclassical model. 

\subsection{Semiclassical Tavis-Cummings model}
The Tavis-Cummings model represents an interaction between a simple harmonic oscillator and a linear top for which there is a
classical model which we now define. We choose the classical model so that the equations of motion are of the same form as
the Heisenberg equations of motion for the quantum model.  The classical Hamiltonian is defined as 
\begin{equation}
{\cal H}=X{\cal J}_x-Y{\cal J}_y+\chi EX
\end{equation}
where $X,Y$ are respectively the canonical oscillator position and momentum variables with the canonical Poisson bracket
$\{X,Y\}=1$, while ${\cal J}_k$ are the three components of angular momentum for a classical top with the canonical Poisson
brackets
$\{{\cal J}_i,{\cal J}_k\}=\sum_{k}\epsilon_{ijk}{\cal J}_k$. The equations of motion for a canonical coordinate $w$ is given
as usual by Poisson bracket with the Hamiltonian $\dot{w}=\{w,H\}$.  The equations of motion are, 
\begin{eqnarray}
\dot{X} & = & -{\cal J}_Y\\
\dot{Y} & = & -{\cal J}_X-\chi\\
\dot{{\cal J}_x} & = & -Y{\cal J}_z\\
\dot{{\cal J}_y} & = & -X{\cal J}_z\\
\dot{{\cal J}_z} & = & X{\cal J}_y+Y{\cal J}_x
\end{eqnarray}
 Note that these equations have a conservation law ${\cal J}_x^2+{\cal J}_y^2+{\cal J}_z^2=\mbox{constant}$.

We now justify this choice of classical Hamiltonian by noting that the Heisenberg equations of motion for the Hamiltonian
Eq(\ref{drivenTC}) have the same form as the semiclassical equations of motion with all variables replaced by the corresponding
operators. We thus see that the semiclassical equations result form taking moments of the Heisenberg equations and factorising
all product moments. 
 The factorisation assumptions ignores correlations which scale as $1/N$ for the scaled
operators $\hat{J}_\sigma/N$.  The conservation law ${\cal J}_x^2+{\cal J}_y^2+{\cal J}_z=constant$ is a reflection of the
operator relation
\begin{equation}
\hat{J}^2=\frac{N}{2}\left (\frac{N}{2}+1\right )
\end{equation}
which in the semiclassical limit indicates that ${\cal J}_x^2+{\cal J}_y^2+{\cal J}_z=\frac{N^2}{4}$.

The classical equations have one nontrivial fixed point at $X^*=Y^*={\cal J}_y^*=0$ and  
${\cal J}_x^*=-\chi$, $ {\cal J}_z^*=\sqrt{N^2/4-\chi^2}$.
However as the conservation law requires that $|{\cal J}_x|\leq N/2$ we see that we must have
\begin{equation}
\frac{2E}{N\Omega}\leq 1\ \ \ \  \mbox{(below threshold)}
\end{equation}
which corresponds to an energy of ${\cal H}=0$.
We will refer to this as the {\em below threshold} case.
As $E$ is increased from zero, the fixed point for the angular momentum system rotates about the ${\cal J}_y$ direction
eventually reaching the equatorial plane at ${\cal J}_x=-N/2$ at the threshold condition. The oscillator system always
has zero amplitude below threshold.  If we linearise around this fixed point we discover that it is an unstable
hyperbolic point with time constant proportional to
$\frac{1}{\sqrt{{\cal J}_z^*}}$. Note that this time constant goes to infinity as the fixed point is approached as is typical
for a hyperbolic fixed point. 

We now consider the {\em above threshold} case
\begin{equation}
\frac{2E}{N\Omega}\geq 1\ \ \ \  \mbox{(above threshold)}
\end{equation}
Clearly the value of $|{\cal J}_x|$ cannot increase above $N/2$. 
 Indeed there is no fixed point above threshold. However there is a special solution curve that continuously joins to the
below threshold case for phase curves with ${\cal H}=0$.

\begin{figure}
\center{\epsfig{figure=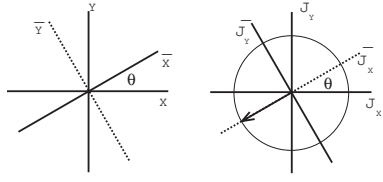,width=50mm}}
\caption{An illustration of the canonical transformation used in the semiclassical equations above threshold} 
\label{fig_one}
\end{figure}

 To see this we consider making a canonical transformation  by a rotation in both the $X-Y$ plane and in the
${\cal J}_x,{\cal J}_y$ plane (see figure \ref{fig_one}). The canonical transformations are
\begin{eqnarray}
X & = & \bar{X}\cos\theta+\bar{Y}\sin\theta\\
Y  & = & \bar{Y}\cos\theta-\bar{X}\sin\theta\\
{\cal J}_x & = & \bar{{\cal J}}_x\cos\theta-\bar{{\cal J}}_y\sin\theta\\
 {\cal J}_y & = & \bar{{\cal J}}_x\cos\theta+\bar{{\cal J}}_y\sin\theta
\end{eqnarray}
The Hamiltonian then takes the form
\begin{equation}
{\cal H}=\bar{X}(\bar{{\cal J}}_x+\chi\cos\theta)-\bar{Y}(\bar{{\cal J}}_y-\chi\sin\theta)
\label{sc_ham}
\end{equation}

The phase curves with ${\cal H}=0$ now correspond to either
\begin{eqnarray}
\bar{X} & = & 0\ \ \ \ ;\ \ \ \bar{{\cal J}}_y  =  \chi\sin\theta
\end{eqnarray}
or
\begin{eqnarray}
\bar{Y} & = & 0\ \ \ \ ;\ \ \ \bar{{\cal J}}_x  =  -\chi\cos\theta
\end{eqnarray}
These phase curves smoothly join the fixed point at threshold if ${\cal J}_x=-N/2$ which implies 
\begin{equation}
\cos\theta=\frac{N\Omega}{2E}
\end{equation}
These solutions are illustrated in figure \ref{fig_one}. Note that as $E\rightarrow \infty$ we have that $\bar{{\cal J}}_x$
eventually points in the direction of $-{\cal J}_y$ while phase curve in the oscillator phase space points along the $Y$ axis,
indicating that for large  driving the system is essentially a particle in a linear potential which accelerates at constant
rate. These results were first obtained by Alsing and Carmichael\cite{AlsingCar}.

\section{Quantum states}
First note that the ground state when there is no driving is $|j,-j\rangle_z\otimes|0\rangle_v$ with a zero eigenvalue. This
ground state corresponds to the fixed point of the semiclassical model with zero oscillator amplitude and angular momentum
pointing in the $-{\cal J}_z$ direction.   We postulate that as the driving is increased form zero the ground state of the
Hamiltonian Eq(\ref{drivenTC}) is given by 
\begin{equation}
|{\cal E}_0\rangle= S(r)R(\theta)|j,-j\rangle_z\otimes|0\rangle_v
\end{equation}
where $ |j,-j\rangle_z\otimes|0\rangle_v$ corresponds to all ions in the ground state and the vibrational mode in the ground
state. The operator $S(r)$ is a squeezing operator defined by 
\begin{equation}
S^\dagger(r)aS(r)=\mu a+\nu a^\dagger
\end{equation}
with $\mu=\cosh r,\ \ \nu=\sinh r$. 

The rotation operator $R(\theta)$ is defined by 
\begin{equation}
R(\theta)=e^{-\theta(\hat{J}_+-\hat{J}_-)}
\end{equation}
and corresponds to a rotation of $2\theta$ around the $\hat{J}_y$ axis. 
Consider now
\begin{equation}
H_I|{\cal E}_0\rangle=SR\left (R^\dagger S^\dagger H SR\right )|j,-j\rangle_z\otimes|0\rangle_v
\end{equation}
If we now transform the Hamiltonian and require that
\begin{equation}
R^\dagger S^\dagger H SR|j,-j\rangle_z\otimes|0\rangle_v=0
\end{equation}
we find the following conditions,
\begin{eqnarray}
\nu(1+\cos2\theta) & = & \mu(1-\cos2\theta)\\
\Omega j\sin2\theta & = &  E
\end{eqnarray}
which requires that 
\begin{equation}
\cos2\theta=e^{-2r}
\end{equation}
and the ground state energy is taken to be ${\cal E}_0=0$. The ground state is thus a product of a squeezed state for the
vibrational mode and a rotated angular momentum state, rotated about the $\hat{J}_y$ axis. 

The above results are consistent with the semiclassical approximation. The mean amplitude of a
squeezed vacuum state is zero, corresponding to the semiclassical 
fixed point at $\bar{X}=\bar{Y}=0$ while the rotation around
the
$\hat{J}_y$ axis corresponds to the semiclassical fixed point at $\bar{{\cal J}}_x=-\chi$.

If we continue to increase $E$ above the threshold value the system adiabatically follows a zero energy state, although this is
no longer a ground state. In fact the canonical transformation used in the semiclassical analysis can be applied to the quantum
operator valued Hamiltonian. The result is the same as the semiclassical case, Eq (\ref{sc_ham}) with all variables replaced
with the corresponding operators. The zero energy state then corresponds to the zero energy eigenstate of
$\hat{Y}\cos\theta-\hat{X}\sin\theta$ with
$\cos\theta=N\Omega/2E$. This is of course just a rotated, infinitely squeezed state. The electronic state is likewise a
 angular momentum eigenstate rotated from $|j,-j\rangle$ in the equatorial plane (orthogonal to $\hat{J}_z$). Thus above
threshold the zero energy eigenstate deforms continuously from the state at threshold.

\begin{figure}
\center{\epsfig{figure=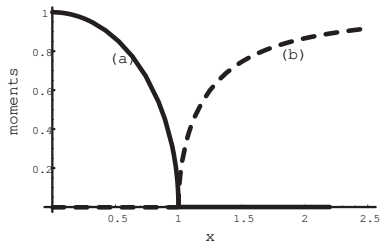,width=50mm}}
\caption[]{The scaled monents, (a) $2|\langle J_x\rangle|/N$ , and (b) $ 2|\langle J_x\rangle|/N$ plotted versus the scaled
driving strength $x=2E/(N\Omega)$. } 
\label{fig_two}
\end{figure}

Let us summarise these results. For no driving the ground state corresponds to the oscillator in the ground state and all ions
in the ground state. As the driving is increased, but kept below threshold,  this state deforms to a squeezed oscillator state
while the collective spin system begins to rotate about the $\hat{J}_y$ axis. 
Note that the mean oscillator amplitude $\langle a\rangle$ remains zero as does the mean of the $y$-component of the collective
spin. As the driving increases through the threshold value, this state changes its character so that a non zero value of
$\hat{J}_y$ is acquired and the oscillator is infinitely squeezed in a direction at an angle $\cos\theta=N\Omega/2E$ to the
below threshold squeezing. This morphological change of the state as the driving passes the semiclassical critical point is a
quantum phase transition. The quantum phase transition can be seen in the mean value for $\hat{J}_y$ and $\hat{J}_z$ as shown
in figure \ref{fig_two}. 
Below threshold the scaled mean values are given by
\begin{eqnarray}
\frac{\langle \hat{J}_y\rangle}{N/2} & = & 0\\
\frac{\langle \hat{J}_z\rangle}{N/2} & = & -\sqrt{1-x^2}
\end{eqnarray}
and above threshold we have
\begin{eqnarray}
\frac{\langle \hat{J}_y\rangle}{N/2} & = & -\sqrt{1-\frac{1}{x^2}}\\
\frac{\langle \hat{J}_z\rangle}{N/2} & = & 0
\end{eqnarray} 
where $x=2E/N\Omega$. 

What are the experimental manifestations of this transition ? Needless to say no one is ever going to observe an infinitely
squeezed state in an experiment. So  what does  happens at
$2\theta=\pi/2$  when the electronic state is the
$\hat{J}_x$ eigenstate
$|j-j\rangle_x$ and the vibrational mode appears to be infinitely squeezed ? Is such a state physically possible ? Suppose for
example we begin in the ground state of the Hamiltonian with no driving ($E=0$) which is simply 
$|j,-j\rangle_z\otimes|0\rangle_v$, and adiabatically increase the driving strength. It would appear that the system would then
adiabatically evolve into the squeezed vibrational state described above. If we were ever able to reach the case
$2\theta=\pi/2$ we would have reached an infinite energy state for the vibrational mode at a finite driving strength. Clearly
this is not possible and to understand why it is useful to reconsider the semiclassical dynamics for this model. The adiabatic
approximation requires that we vary the driving strength on a time scale slower than all other time scales in the system. The
key time scale for the ground state variation is just the time scale associated with the hyperbolic unstable fixed point,
$(N^2/4-\chi^2)^{-1/2}$, which goes to infinity as we approach 
$\frac{2E}{\Omega N}=1$. Thus the adiabatic increase of the driving must proceed infinitely slowly, that is it must be switched
to the finite value $E=\frac{\Omega N}{2}$ in an infinite amount of time. This pumps an infinite amount of energy into the
system and results in infinite squeezing in the centre of mass vibrational mode. Obviously in practice this cannot be achieved
so the totally squeezed ground state is not possible. However it will still be possible to achieve some squeezing of the
vibrational mode at smaller values of the driving. This would make an interesting observation for current ion trap experiments
even with only a few ions. The squeezing of the vibrational mode can be observed using the dynamical method of
reference \cite{Wineland-squeeze}

In current ion trap experiments, laser cooling techniques allow the centre of mass mode to be prepared in the ground state.
Unfortunately it does not stay there. Heating due to a variety of sources, including fluctuating linear potentials, lead to
an irreversible evolution away from the ground state. If such heating is present during the coupling of the electronic and
vibrational motions, irreversible dynamics will be spread to the collective spin degrees of freedom as well. 

As an example we
consider what happens if we use the Tavis-Cummings interaction (excitation on first red sideband) in the presence of strong
heating. Heating of the centre-of-mass mode due to fluctuating liner potentials may be described in the interaction picture by
the master equation,
\begin{equation}
\frac{dW}{dt}=-i\Omega[a\hat{J}_++a^\dagger\hat{J}_-,W]+\frac{\gamma}{2}\left ({\cal D}[a]+{\cal D}[a^\dagger]\right )W
\end{equation}
where $W$ is the density operator for the spin and vibrational degrees of freedom and the superoperator ${\cal D}$ is defined
by  $
{\cal D}[A]\rho=2A\rho A^\dagger-A^\dagger A\rho-\rho A^\dagger A\ \ .
$
The irreversible term corresponds to two point processes in which phonons are removed or added from centre of mass mode at
the rates $\gamma\langle a^\dagger a\rangle$ and $\gamma\langle a a^\dagger\rangle$  respectively. This  does not change
any first order moments, however it does lead to a diffusion in energy as
$
\frac{d\langle a^\dagger a\rangle}{dt} =\gamma\ \ .
$
The effect of heating can be included in the semiclassical analysis by adding an appropriate stochastic term. In the Ito
calculus\cite{Gardiner} the effect is to add to the equations for $X,Y$ terms of the form
\begin{eqnarray}
dX & = & (\ldots)+\sqrt{\gamma}dW_x(t)\\
dY & = & (\ldots)+\sqrt{\gamma}dW_y(t)
\end{eqnarray}
where $dW_i(t)$ are independent Wiener processes. If the heating rate is small enough these terms can be neglected. However if
they are large new steady states can occur in the semiclassical and quantum descriptions which will be described in a future
publication. 

We would like to thank Howard Carmichael for useful discussions.

\end{document}